\documentclass[journal,twoside]{IEEEtran}

\usepackage[pdftex]{graphicx}
\usepackage[cmex10]{amsmath}

\usepackage{pifont}
\usepackage{amssymb}
\usepackage{textcomp}
\usepackage{multirow}
\usepackage{color}
\usepackage{cite}

\begin{document}
\title{Probabilistic Modeling and Simulation of Transmission Line Temperatures under Fluctuating Power Flows}

\author{Markus~Schl\"apfer,~\IEEEmembership{Student Member,~IEEE},
        and~Pierluigi~Mancarella,~\IEEEmembership{Member,~IEEE}%      
\thanks{M. Schl\"apfer is with the Laboratory for Safety Analysis, Institute of Energy Technology, ETH Zurich, 8092 Zurich, Switzerland.}% <-this % stops a space
\thanks{P. Mancarella is with the Control and Power Group, Department of Electrical and Electronic Engineering, Imperial College, London SW7 2AZ, UK.}% <-this % stops a space
%\thanks{Manuscript received April 19, 2005; revised January 11, 2007.}
}

\maketitle
\begin{abstract}
Increasing shares of fluctuating renewable energy sources induce higher and higher power flow variability at the transmission level. The question arises as to what extent existing networks can absorb additional fluctuating power injection without exceeding thermal limits. At the same time, the resulting power flow characteristics call for revisiting classical approaches to line temperature prediction. This paper presents a probabilistic modeling and simulation methodology for estimating the occurrence of critical line temperatures in the presence of fluctuating power flows. Cumbersome integration of the dynamic thermal equations at each Monte Carlo simulation trial is sped up by a specific algorithm that makes use of a variance reduction technique adapted from the telecommunications field. The substantial reduction in computational time allows estimations closer to real time, relevant to short-term operational assessments. A case study performed on a single line model provides fundamental insights into the probability of hitting critical line temperatures under given power flow fluctuations. A transmission system application shows how the proposed method can be used for a fast yet accurate operational assessment.
\end{abstract}
\begin{IEEEkeywords}
Line temperature dynamics, fluctuating power generation, Monte Carlo simulation, variance reduction technique.
\end{IEEEkeywords}

%\IEEEpeerreviewmaketitle

\section*{Nomenclature}

The following listing contains only the main symbols as used in the paper. Other symbols are defined in the text, when required.
\vspace{-0.1cm}
\subsection{Parameters}

\begin{flushleft}
\begin{IEEEdescription}[\IEEEusemathlabelsep]

\item[$C$] Number of aggregated generating units.
\item[$m$] Total number of temperature thresholds.
\item[$N$] Number of main trials.
\item[$n_i$] Number of retrials starting at $T^{i-1}$.
\item[$A_\ell^c(t)$] Convected heat loss coefficient of line~$\ell$ [Wm$^{-1}$K$^{-1}$].
\item[$A_\ell^r(t)$] Radiated heat loss coefficient of line $\ell$ [Wm$^{-1}$K$^{-4}$].
\item[$f_g$] State transition frequency of generator $g$ [$s^{-1}$].
\item[$p_{ij}(\nu)$] State transition probability.
\item[$\textbf{P}(\nu)$] Transition probability matrix.
\item[$P^s_i$] Stationary probability of state $i$.
\item[$Q_\ell^s(t)$] Solar heat gain for line $\ell$ [W$\,$m$^{-1}$].
\item[$r_\ell^{ref}$] Reference resistance of line $\ell$ at $T_\ell^{ref}$ [p.u.].
\item[$S_B$] Base apparent power [MVA].
\item[$T^i$] Temperature threshold $i$ [K].
\item[$T_\ell^{a}$] Ambient temperature of line $\ell$ [K].
\item[$T_\ell^{r}$] Maximum allowed operating temperature of line $\ell$ [K].
\item[$T_\ell^{ref}$] Reference temperature of line $\ell$ [K].
\item[$t_0$] Starting time of the analysis period [s].
\item[$t_e$] Stopping time of the analysis period [s].
\item[$x_\ell$] Reactance of line $\ell$ [p.u.].
\item[$\alpha_\ell^{ref}$] Thermal resistivity coefficient of line $\ell$ [K$^{-1}$].
\item[$\epsilon$] Accuracy level.
\item[$\lambda_g$] Transition rate from up to down state of generator $g$ [s$^{-1}$].
\item[$\mu_g$] Transition rate from down to up state of generator $g$ [s$^{-1}$].
\item[$\nu$] State sampling frequency [s$^{-1}$].
\item[$\rho_\ell$] Heat capacity of line $\ell$ [J$\,$m$^{-1}$$\,$K$^{-1}$].
\item[$\tau_i$] Mean holding time in state $i$ [s]. 
\item[$\underline{\omega}_\ell(t)$] Vector of meteorological parameters for line $\ell$.
\end{IEEEdescription}
\end{flushleft}
%
%\vspace{-0.1cm}
\subsection{Functions}
%\vspace{-0.1cm}
\begin{IEEEdescription}[\IEEEusemathlabelsep]
\item[$E(\bullet)$] Expectation.
\item[$F(\bullet)$] Cumulative distribution function.
\item[$I_{\ell,p}(\bullet)$] Phase current on line $\ell$ [A].
\item[$I_{\ell,s}(\bullet)$] Ampacity of line $\ell$ under steady state conditions  [A].
\item[$Q_\ell^c(\bullet)$] Convected heat loss of line $\ell$ [W$\,$m$^{-1}$].
\item[$Q_\ell^j(\bullet)$] Ohmic loss of line $\ell$ phase conductor [W$\,$m$^{-1}$].
\item[$Q_\ell^r(\bullet)$] Radiated heat loss of line $\ell$ [W$\,$m$^{-1}$].
\item[$r_\ell(\bullet)$] Resistance of line $\ell$ [p.u.].
\item[$RE(\bullet)$] Relative error.
\item[$Var(\bullet)$] Variance.
\end{IEEEdescription}
%
%\vspace{-0.1cm}
\subsection{Variables}
%
%\vspace{-0.1cm}
\begin{IEEEdescription}[\IEEEusemathlabelsep]\IEEEsetlabelwidth{3cm}
\item[$p_i$] Conditional probability that $T_\ell(t)$ reaches $T^i$, given that it has already reached $T^{i-1}$.
\item[$\underline{s}(t)$] System state vector.
\item[$T_\ell(t)$] Temperature of line $\ell$ [K].
\item[$V_i(t)$] Voltage magnitude at node $i$ [p.u].
\item[$\gamma$] Probability that $T_\ell(t)$ reaches $T^m$.
\item[$\hat{\gamma}$] Estimator of $\gamma$ in all trials.
\item[$\hat{\gamma}_k$] Estimator of $\gamma$ in trial $k$.
\item[$\theta_i(t)$] Voltage angle at node $i$ [rad].
\item[$\chi_{a,k}$] Number of events in trial $k$ at which $T_\ell(t) \geq T^m$.
\end{IEEEdescription}
%
%\newpage
%
\section{Introduction}
%\vspace{-0.1cm}
\IEEEPARstart{I}{ncreasing} volumes of fluctuating renewable energy sources, as exemplified by wind energy conversion, are leading to more variable and less predictable power flows in networks \cite{Smith:2007}. This also implies a decrease in the average network utilization and, possibly, an increase in the probability of hitting thermal limits due to peak flows. While on the long run network assets will eventually be upgraded, the expansion of the transmission system is rather slow with time horizons up to several years. The existing assets therefore need to be used as efficiently as possible. As a consequence, the question arises whether present classical concepts for estimating the loading capability of overhead lines are adequate to cope with these rapid developments. In this respect, it has recently been shown that in order to fulfill the sag clearance requirements, the direct use of temperature rather than power limits allows for a significantly more precise and less conservative loadability assessment \cite{Banakar:2005, Alguacil:2005}. New approaches and tools are crucial not only for contingency analysis, but also for releasing available power transfer capability, potentially underestimated by classical line rating methodologies, to increase the amount of fluctuating renewable energy sources that can be securely integrated in the system as well as to increase the volume of energy that can be traded between nodes. The uncertainty given by the stochastic nature of renewable sources thereby calls for approaches based on probability concepts \cite{Billinton:1996, Schlapfer:2012}. In this outlook, existing probabilistic models for different renewable energy sources (e.g., \cite{Sayas:1996, Papaefthymiou:2009,Silva:2010}), theoretical advances in simulation speed-up techniques (e.g., \cite{Villen-Altamirano:2006, daSilva:2010}) and the evolution of computational power pave the way to Monte Carlo-based methodologies.

On these premises, the present paper introduces a probabilistic modeling and simulation approach for assessing the impact of power flow fluctuations on the occurrence probability of maximum allowed transmission line temperatures. The time-varying temperatures are explicitly modeled and calculated from the heat balance equations. Coupling to and interaction with the AC power flow variables is carried out through the ohmic losses and the temperature dependent conductor resistances. Monte Carlo simulations are used to generate the probabilistic information on the line temperature dynamics in the presence of uncertainties adherent to, for instance, fluctuating wind turbine generation or forced outages of conventional generating units. The proposed methodology can be applied regardless of the specific probabilistic model used for capturing the relevant stochastic phenomena. In order to overcome the downside of the slow simulation speed when performing massive Monte Carlo extractions with continuous integration of the heat balance equations, a specific algorithm has been developed for the problem under analysis. It is based on a technique for the fast simulation of rare events called RESTART (REpetitive Simulation Trials After Reaching Thresholds), mainly adopted hitherto in the telecommunications field.

The paper is organized as follows. Section II introduces the electrothermal model for the dynamic calculation of overhead line temperatures. Section III discusses the modeling framework and the Monte Carlo accelerated algorithm for the probabilistic line temperature assessment. Section IV reports the simulation results from a single line example, revealing the impact of different flow fluctuation characteristics on the line temperature dynamics, as well as demonstrating the efficiency of the acceleration algorithm. An additional case study is carried out for a transmission network with fluctuating wind power injections and generator failure events, conventionally modeled through Markov chains, highlighting the benefits of the methodology for short-term operational purposes. \mbox{Section V} concludes the work.
%
%\vspace{-0.1cm}
\section{Electrothermal model for line temperature dynamics}\label{sec:temp}
%\vspace{-0.1cm}
%
The electrothermal model described here aims at calculating the time-varying transmission line temperatures as driven by the power flow fluctuations and meteorological conditions. Each conductor of a transmission line $\ell$, connecting node $y$ with node $z$, is heated by its temperature dependent ohmic losses $Q_\ell^j(T_\ell(t))=I_{\ell,p}^2(t)R(T_\ell(t))$, with $I_{\ell,p}(t)$ being the phase current, and by the solar heat gain $Q_\ell^{s}(t)$  \cite{IEEE:2006}. Convection $Q^{c}_\ell(T_\ell(t))$ and radiation $Q^{r}_\ell(T_\ell(t))$ are responsible for cooling. This heat balance yields the following differential equation for the conductor temperature $T_\ell(t)$:
\begin{equation}\label{eq:HBE}
\rho_{\ell}\frac{d}{d t} T_{\ell}(t)= Q_\ell^j(T_\ell(t))+Q^{s}_\ell(t)-Q^{c}_\ell(T_\ell(t))-Q^{r}_\ell(T_\ell(t))
\end{equation}
where $\rho_{\ell}$ is the heat capacity of the conductor. As all three conductors have the same electrical and thermal characteristics, $T_\ell(t)$ can be regarded as the transmission line temperature. Its time evolution is derived by numerically solving (\ref{eq:HBE}), applying standard integration methods. Following the notation given in \cite{Banakar:2005} the convection and radiation terms can be calculated by
\begin{equation}\label{eq:HBEb}
Q^{c}_\ell(T_\ell(t)) = A^c_\ell(\underline{\omega}_\ell(t))(T_\ell(t)-T^a_\ell(t))\\
\end{equation}
\begin{equation}\label{eq:HBEc}
Q^{r}_\ell(T_\ell(t)) = A^r_\ell(\underline{\omega}_\ell(t))([T_\ell(t)]^4-[T^a_\ell(t)]^4)
\end{equation}
where $A^c_\ell(t)$ and $A^r_\ell(t)$ are the convected and radiated heat loss coefficients, $\underline{\omega}_\ell(t)$ is the vector of weather parameters and $T^a_\ell(t)$ is the ambient temperature. Assuming a $\pi$-equivalent line model with base power $S_B$ and neglecting the shunt conductance, $Q_\ell^j(t)$ is obtained from the three phase, per unit (p.u.) Joule losses $q_\ell^j(t)$ as
\begin{equation}\label{eq:lossA}
Q_\ell^j(t)=q_\ell^j(t) S_B/3
\end{equation}
where $q_\ell^j(t)$ is given by the power flow variables as
%%%%%Equation
{\setlength\arraycolsep{4pt}
\begin{eqnarray}\label{eq:lossB}
q_{\ell}^j(t) & = & \frac{r_{\ell}(t)}{r_{\ell}^2(t)+x_{\ell}^2} \big[V_y^2(t) + V_z^2(t) {} \nonumber\\
\vspace{2cm} &&  -2 V_y(t)V_z(t) \cos(\theta_y(t)-\theta_z(t)) \big].
\end{eqnarray}}
%%%%%Equation
The temperature of a transmission line, in turn, affects its resistance, with a behavior that can be approximated by a linear model \cite{Banakar:2005}
\begin{equation}\label{eq:HBEa}
r_\ell(T_\ell(t))=r^{ref}_\ell[1+\alpha_\ell^{ref}(T_\ell(t)-T_\ell^{ref})]
\end{equation}
where $\alpha_\ell^{ref}$ denotes the thermal resistivity coefficient and $r^{ref}_\ell$ is the resistance at the reference temperature $T_\ell^{ref}$. This dependence of the resistance on the actual temperatures can not be neglected for accurate power flow \mbox{calculations \cite{Bockarjova:2007}}. However, after a change of the power flow the voltage magnitudes and angles show only small variations during the resulting temperature transients, having time spans typically in the range of 30 min \cite{Banakar:2005}. This allows for a temporary decoupling of the power flow variables from the actual line temperature, meaning that the resistances need to be updated only periodically after defined time steps or after significant temperature changes. The specific updating rules are to be defined on the basis of each given study case. The variation of the line reactance $x_\ell$ with the temperature is small and can be neglected \cite{Banakar:2005}.

In practice, the operation of transmission lines is usually constrained by thermal ratings, whereas the maximum allowed operating temperatures become converted into ampacities (i.e., maximum current carrying capacities). The ampacity $I_{\ell,s}$ of a transmission line is conventionally calculated on the basis of steady-state thermal ratings \cite{Deb:1968}
\begin{equation}\label{eq:steadyState}
I_{\ell,s}=\sqrt{\frac{Q_\ell^c(T_{\ell}^r)+Q_\ell^r(T_{\ell}^r)-Q_\ell^s}{R(T_{\ell}^r)}}
\end{equation}
where $T_{\ell}^r$ is the maximum allowed operating temperature of the line. Determining $I_{\ell,s}$ can be either based on conservative assumptions for the weather parameters, or on the actual conditions \cite{Deb:1968,517499,Seppa:2000}. The latter usually allows for a higher ampacity but requires the monitoring of the line temperatures and meteorological data. 
\section{Modeling framework for line temperature probabilistic assessment}
\label{sec:frame}
\subsection{Conceptual Basics}
In the presence of fluctuating power flows, the electrothermal model can be deployed in Monte Carlo-based simulations for estimating the probability that a transmission line reaches a certain temperature within a given time span. Depending on the specific study, the flow fluctuations can be induced by deterministic and probabilistic models for various phenomena, ranging from time-varying demand patterns and short-term energy trading behavior to forced component outages and, essentially, fluctuating renewable energy sources. In order to overcome the prohibitive downside of the slow simulation speed coming along with the extensive simultaneous solution of both the power flow and the heat balance equations, a specific variance reduction algorithm has been developed. The algorithm borrows from the RESTART technique for the fast simulation of rare events, which has been proposed in the telecommunications field. 

The line temperature assessment methodology introduced here is completely general, whereas the RESTART technique can be applied to both Markovian and non-Markovian processes \cite{Villen-Altamirano:2006}. This allows combining the methodology with a broad spectrum of different probabilistic models. Examples are Markov models for the stochastic failure and repair behavior of generators, transmission lines and other components \cite{Billinton:1994}, or ARIMA models for the power fluctuations from wind farms \cite{Milligan:2003, ANL:2009}. The methodology is detailed below. 
%\vspace{-0.1cm}
%
\subsection{Accelerated Monte Carlo Simulation for Dynamic Line Temperature Estimation}
%\vspace{-0.1cm}
%
The objective of the assessment is to estimate the probability $\gamma$ that the temperature $T_\ell(t)$ of a transmission line $\ell$ reaches the maximum allowed operating temperature $T_\ell^r$ within a time period $[t_0,t_e)$. A sequential Monte Carlo simulation therefore samples the chronological state transitions of each relevant system component \cite{Billinton:1994}. An example is the time-varying power output state of a wind farm. The state of the overall power system at each time is then given by the combination of all respective component states. The chronological system state transition process, in turn, is needed for determining the time-varying power flow which eventually allows calculating the dynamics of the transmission line temperatures according to (\ref{eq:HBE}).
A crude Monte Carlo method repeats these simulation steps $N$ times within $[t_0,t_e)$ and estimates the probability ${\gamma}$ as 
\begin{equation}\label{eq:gammaCrude}
\gamma \approx \hat{\gamma}=\frac{1}{N}\sum_{i=1}^N \chi_{c,i}
\end{equation}     
with $\chi_{c,i}$ being the zero-one indicator that $T_\ell(t)$ reaches $T_\ell^r$ within trial $i$. The accuracy of the estimation can be quantified by its relative error
%%%%%Equation%%%%%
\begin{equation}\label{eq:crudeREp}
RE(\hat{\gamma}):= \frac{\sqrt{Var(\hat{\gamma})}}{E(\hat{\gamma})} = \sqrt{\frac{1-\hat{\gamma}}{N\hat{\gamma}}}. 
\end{equation}
%%%%%Equation%%%%%

\begin{figure}[!t]
\centering
\includegraphics[width=3.5in]{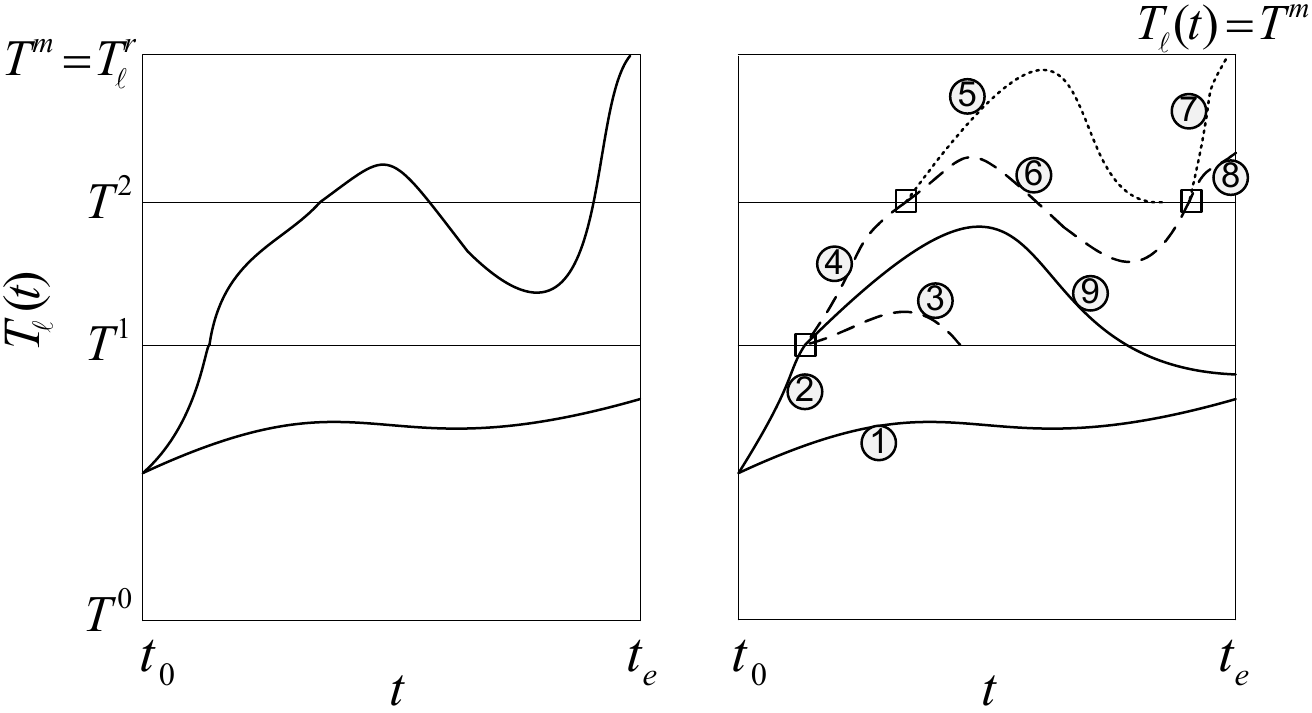}
\caption{Left: crude simulation, right: simulation by means of RESTART with $N=2$, $m=3$, $n_1=1$, $n_2=3$ and $n_3=2$. The numbers correspond to an exemplary sequence of consecutive simulation steps (see main text). The squares indicate the initial states for the retrials.}\label{Restart}
\end{figure}

The basic idea of RESTART is to perform a higher number of simulation trials in those regions of the state space, where the event of interest is more often provoked \cite{Villen-Altamirano:2006, Garvels:1998}. Opposite to other variance reduction methods for Monte Carlo simulations such as the importance sampling technique \cite{Billinton:1994}, RESTART has no influence on the sequence of the stochastic events in absolute time. In the following we introduce the adaptation of the technique to the line temperature estimation problem. Let us first divide the temperature state space $[T^0,T_\ell^{r}]$ into $m$ intermediate intervals $[T^0,T^1),[T^1,T^2),\ldots,[T^{m-1},T^m]$ with thresholds $T^0<T^1<\ldots<T^m=T_\ell^{r}$. Starting at $t_0$ with $T^0 \leq T_\ell(t_0)<T^1$, the line temperature has to pass all intervals in order to reach $T^m$. We further denote as $p_i$ the conditional probability that $T_\ell(t)$ reaches threshold $T^i$ before the time reaches $t_e$, given that $T_\ell(t)$ has already passed threshold $T^{i-1}$. The occurrence probability of reaching $T^m$ then becomes
%%%%%Equation%%%%%%
\begin{equation}\label{eq:RESTARTGammaI}
\gamma= p_1p_2\cdots p_m.
\end{equation}
%%%%%Equation%%%%%%
A crude Monte Carlo method repeatedly simulates the system within $[t_0,t_e)$ (see Fig. \ref{Restart}, left). The higher a threshold $T^i$ the less sample paths are reaching it and the less accurate is the estimation of $p_i$. In order to compensate for this lack of trials in the regions closer to $T^m$, RESTART stores the system state as soon as $T_\ell(t)$ reaches a threshold $T^{i}$, and splits the sample path into $n_{i+1}$ retrials for the time during which it stays above this threshold (see Fig. \ref{Restart}, right). The first $n_{i+1}-1$ paths are stopped when they again fall below $T^i$ in order to avoid simulation time in the regions away from $T^m$. Only the last path is permitted to proceed so that it becomes a continuation of the original path. Consequently, a larger number of trials for accurately estimating each $p_i$ is achieved.
%%%%%Figure%%%%%
%%%%%Figure%%%%%

Different implementation schemes of the RESTART technique have been proposed \cite{Villen-Altamirano:2006, Garvels:1998}. We apply the `global-step' approach for estimating $\gamma$. Its main advantage is the need to store at most $m$ system states only, as indicated by the squares in Fig. \ref{Restart}. Thereby, the unbiased estimator of $\gamma$ is given by
\begin{equation}\label{eq:RESTARTgamma}
\hat{\gamma}= \sum_{k=1}^N \chi_{a,k} \bigg[N \prod_{i=1}^m n_i \bigg]^{-1}
\end{equation}
where $N$ is the total number of main trials starting at $t_0$ and $\chi_{a,k}$ counts how many times $T_\ell(t)$ reaches $T^m$ in trial $k$. The simulation is stopped when the relative error of $\hat{\gamma}$ becomes smaller than a predefined accuracy level $\epsilon$ 
\begin{equation}\label{eq:RESTARTREp}
RE(\hat{\gamma}) \approx \frac{ \sqrt{\sum_{k=1}^N \hat{\gamma}_k^2-N\hat{\gamma}^2}}{N\hat{\gamma}}< \epsilon
\end{equation}
where $\hat{\gamma}_k=\chi_{a,k}/\prod_i n_i$ is the estimate in trial $k$. To maximize the computational gain, the thresholds are chosen in such a way that $p_i$ and $n_i$ reach their quasi-optimal values \cite{Villen-Altamirano:2006}
\begin{equation}\label{eq:optimal}
p_i=e^{-2}; \qquad n_i=\sqrt{1/(p_ip_{i+1})}.
\end{equation}
The position of the thresholds and the values of $n_i$ are determined by performing a pilot run. 

\begin{figure}[!t]
\centering
\includegraphics[width=3.8in]{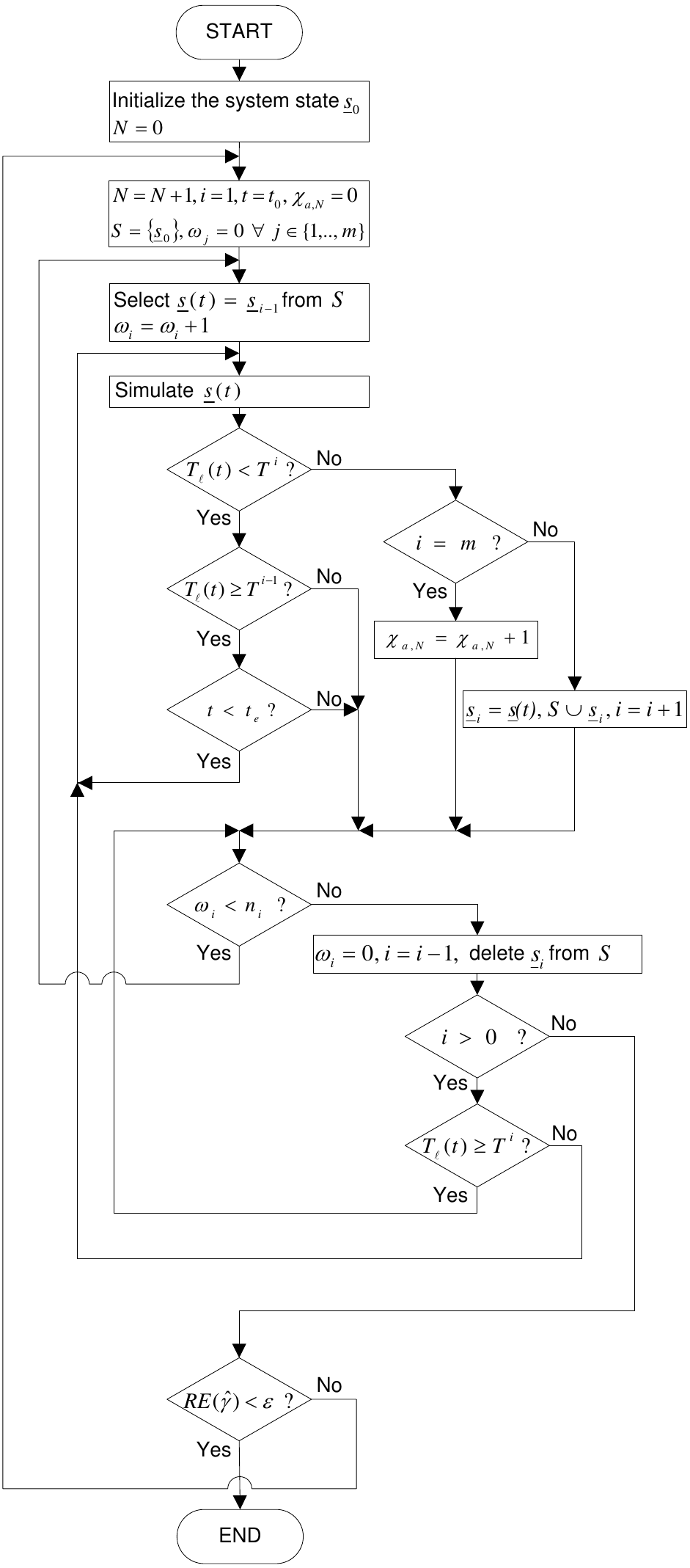}
\caption{Flowchart of the accelerated algorithm for estimating the probability of line temperature $T^m$.}  
\label{flowchart}
\end{figure}

As shown by the flowchart in Fig. \ref{flowchart}, the algorithm consists of several loops. 
%%%%%FIGURE%%%%%
%%%%%FIGURE%%%%%
All variables needed for the calculation of $T_\ell(t)$, such as the actual generator or load states and the actual temperatures of all transmission lines, are stored in the system state vector $\underline{s}(t)$. The time-varying component states are thereby governed by case-specific deterministic and probabilistic models. The outer loop starts the main trial $N$ at $t_0$ with the initial system state $\underline{s}_0$. As soon as this main trial and all triggered retrials have been finished the probability of reaching $T^m$ is estimated, together with the relative error. The inner loops simulate the time evolution of $\underline{s}(t)$ and perform all the consecutive retrials if $T_\ell(t)$ reaches a threshold $T^i$. The heat balance equation is continuously integrated, whereas the line resistances are periodically updated as described in Section \ref{sec:temp}. The numbers in Fig. \ref{Restart}, right, illustrate an exemplary sequence of the resulting simulation steps. The first trial \mbox{(step \ding{192})} does not reach the first temperature threshold and, as in a crude Monte Carlo approach, is only stopped when the simulated time equals $t_e$. When reaching $T^1$ the second trial (step \ding{193}) becomes split at system state $\underline{s}_1$, whereas the first retrial (step \ding{194}) is stopped when falling below this threshold again. The second retrial (step \ding{195}) is again split when reaching $T^2$ at $\underline{s}_2$. Restarting from this stored system state, both retrials (steps \ding{196} and \ding{197}) fall again below $T^2$. As the continuation of the original path only the second retrial (step \ding{197}) is further simulated, reaching again $T^2$. The respective first retrial (step  \ding{198}) eventually reaches $T^m$ and the subsequent second retrial (step \ding{199}) is stopped when the time equals $t_e$. Having thus finished all retrials starting from $T^2$, the algorithm proceeds with the remaining retrial starting from $T^1$ (step  \ding{200}).
\section{Case study applications}
\subsection{Example A: Fluctuating Power Flows on a Single Line}
\begin{figure}[!b]
\centering
\includegraphics[width=3.4in]{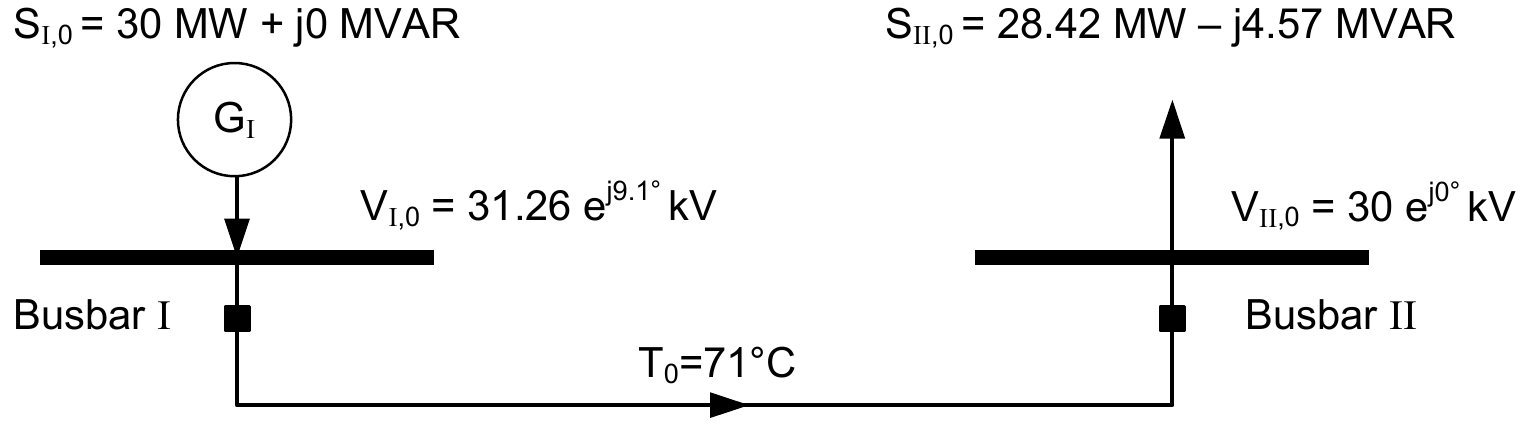}
\caption{Single-line layout}  
\label{single-line}
\end{figure}

\begin{figure}[!b]
\centering
\includegraphics[width=3.3in]{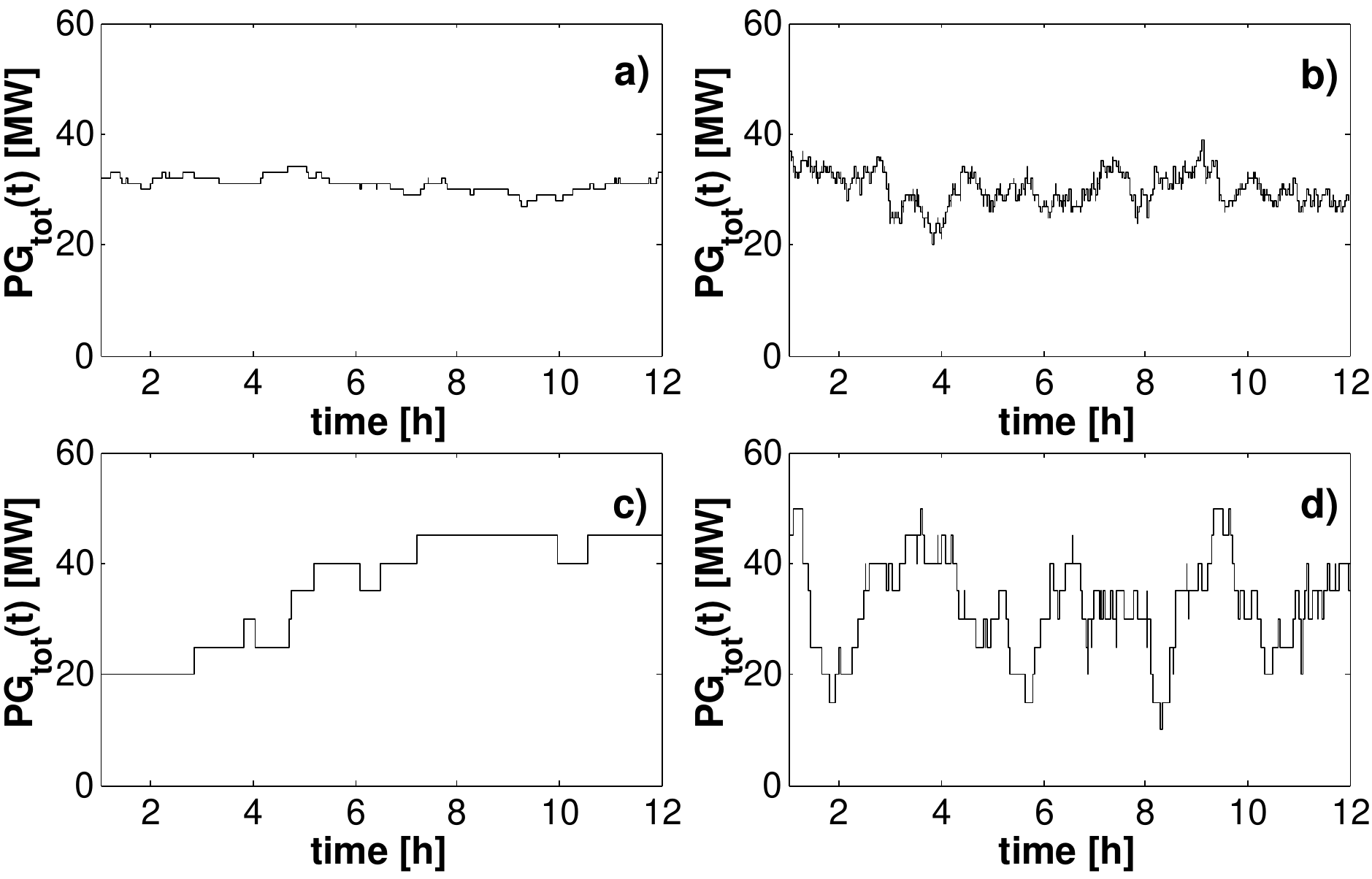}
\caption{Combined power injection patterns. a) $f_g=0.05$h$^{-1}$, $C=1$. b) $f_g=0.5$h$^{-1}$, $C=1$. c) $f_g=0.05$h$^{-1}$, $C=10$ d) $f_g=0.5$h$^{-1}$, $C=10$.}
\label{injection}
\end{figure}

A single line example is used to gain fundamental insights into the impact of different flow fluctuation characteristics on the line temperature dynamics. The values of the power flow variables and line temperature at $t_0=0$ are depicted in \mbox{Fig. \ref{single-line}}. 
%%%%%Figure%%%%%
%%%%%Figure%%%%%
The line consists of three `Drake' 26/7 ACSR conductors with a length of $20$ km. The corresponding parameters for calculating the temperature behavior are given in Table II in Appendix A. The power injection at \mbox{busbar I} is generated by $NG=60$ single units with an individual output of $PG_g^{max}=1$ MW real power and zero reactive power. For the purpose of this experimental study the power ouput $PG_g(t)$ of each single generating unit $g$ is governed by a simple two-state process, whereas $PG_g(t)$ is either $PG_g^{max}$, being in up state, or zero, being in down state. The stochastic alternation between these two states is determined by the transition rates $\lambda_g$ and $\mu_g$, implying exponentially distributed holding times
\begin{equation}\label{eq:genF}
F_g(t_{up}) = 1-e^{-\lambda_g t_{up}} , \qquad F_g(t_{down}) = 1-e^{-\mu_g t_{down}}  
\end{equation}
where $t_{up}$ and $t_{down}$ are the time spans measured from the moment of entering the up state and down state respectively. The state transition frequency is given by
\begin{equation}\label{eq:renewalDensity}
f_g  = \frac{\lambda_g \mu_g}{\lambda_g + \mu_g}
\end{equation}
and corresponds to the average number of up-down-up cycles per time unit. The value of this simple model is the high flexibility to reproduce a large number of combined power output patterns, $PG_{tot}(t)=\sum_{g=1}^{NG}PG_g(t)$, while the mean power output $E(PG_{tot}(t))$ stays constant, allowing to systematically study the impact of different fluctuation characteristics. This is achieved by varying $f_g$ while keeping the ratio $\lambda_g / \mu_g=1$ constant and by aggregating different numbers of generating units into different clusters, within which the units follow simultaneously the same production cycles over time. We denote the size of such a cluster (i.e., the number of aggregated units) as $C$. Comparison between Fig. \ref{injection}, left, and Fig. \ref{injection}, right, shows how an increased value of  $f_g$ leads to a faster fluctuation of the injection at \mbox{busbar I}. According to Fig. \ref{injection}, upper part, compared to Fig. \ref{injection}, lower part, a smaller number of $C$ is leading to smoother time-series, while a large value implies a strong fluctuation around $E(PG_{tot}(t))=30$ MW.
%%%%%Figure%%%%%
%%%%%Figure%%%%%

\begin{figure}[!b]
\centering
\includegraphics[width=3.3in]{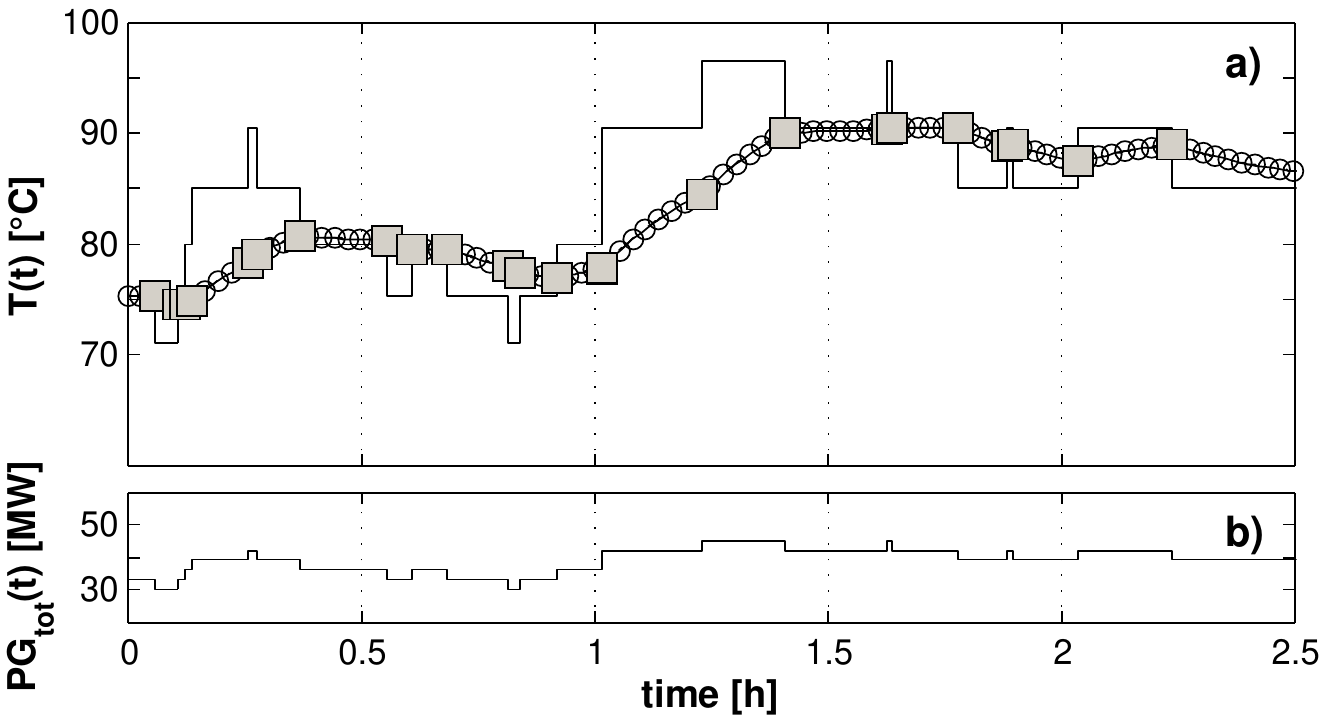}
\caption{Lower part: combined power injection sequence. Upper part: resulting dynamic temperature behavior. The line marked with circles corresponds to the temporary decoupling of the power flow from $T_\ell(t)$, being overlapped by the numerical results without the decoupling (line marked with squares). The line without markers corresponds to the steady-state temperature.}  
\label{tempEvol}
\end{figure}

The algorithm (Fig. \ref{flowchart}) is implemented in Matlab \cite{Matlab:2008}, running on an Intel Xeon E5450 quad-core processor with 2.99 GHz CPU speed. By making use of the temporary decoupling of the power flow variables from the line temperature, the line resistance is updated only in case of a generator output change. Figure \ref{tempEvol} shows an exemplary power injection sequence and the resulting temperature behavior both based on this simplification and without the temporary decoupling. The excellent match confirms the validity of this model assumption. Additionally, Fig. \ref{tempEvol} depicts the steady-state temperatures corresponding to each subsequent power output state. These values differ significantly from the actual temperatures due to the thermal inertia, supporting the need to consider the transient behavior.
%%%%%Figure%%%%%
%%%%%Figure%%%%%

\begin{figure}[!b]
\centering
\includegraphics[width=2.7in]{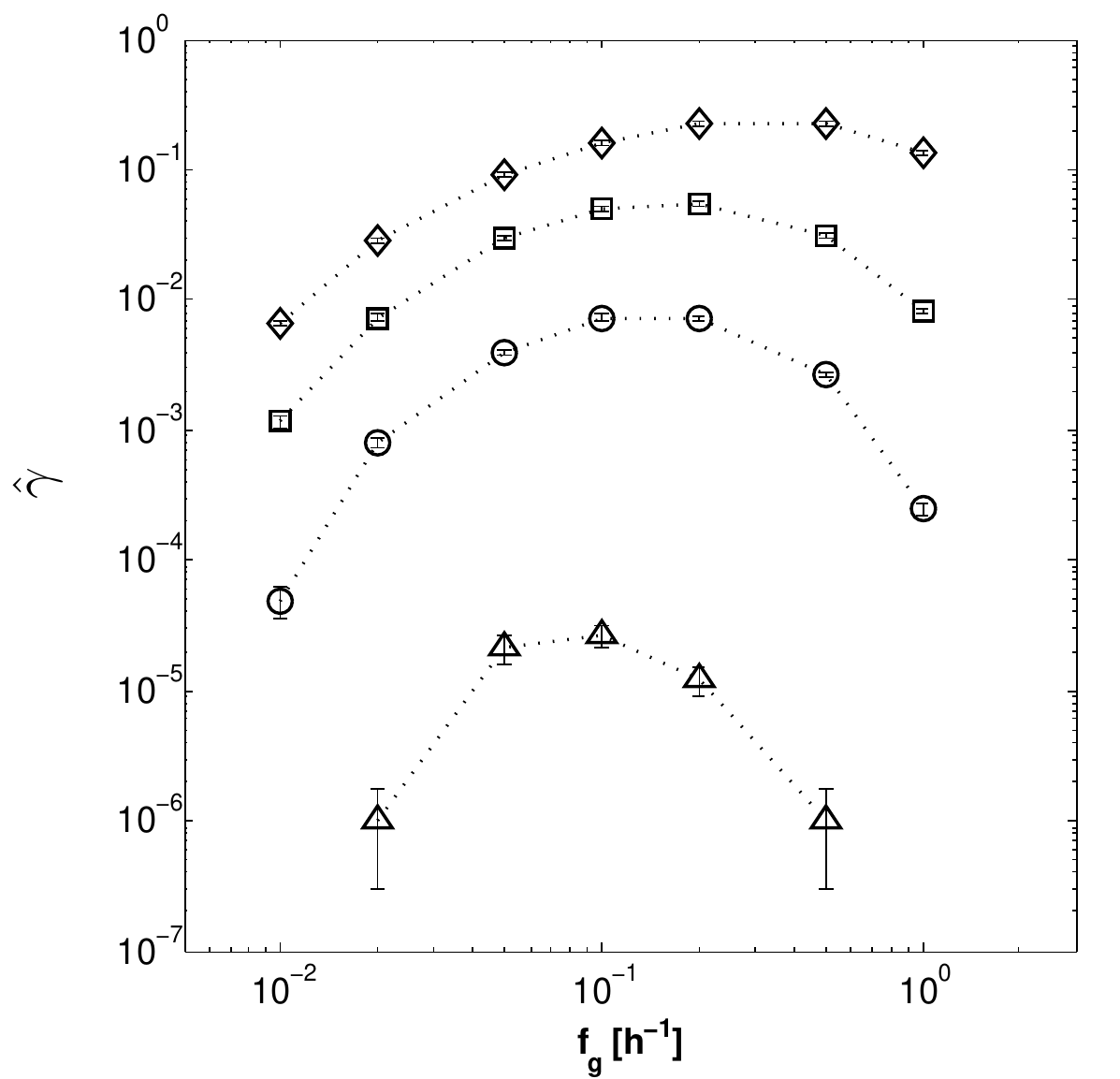}
\caption{Estimates of $\gamma$ in relation to the fluctuation frequency $f_g$ for $C$=1,2,3,5 (triangles, circles, squares and diamonds, respectively). The error bars indicate $RE(\hat{\gamma})$.}
\label{Fig:gamma}
\end{figure}

For this example we define $\gamma$ as the probability that the line temperature reaches the maximum allowed value $T_\ell^r=100\,^{\circ}\mathrm{C}$ within the time interval $[t_0=0$ h, $t_e=12$ h$)$. In order to uncover the effect of the fluctuation, the sensitivity of $\gamma$ with respect to both the `fluctuation frequency' $f_g$ and the `fluctuation magnitude' $C$ has been analyzed. Figure \ref{Fig:gamma} reports the resulting values of $\hat{\gamma}$.
%%%%%Figure%%%%%
%%%%%Figure%%%%%
%%%%%Figure%%%%%
\begin{figure}[!b]
\centering
\includegraphics[width=3.2in]{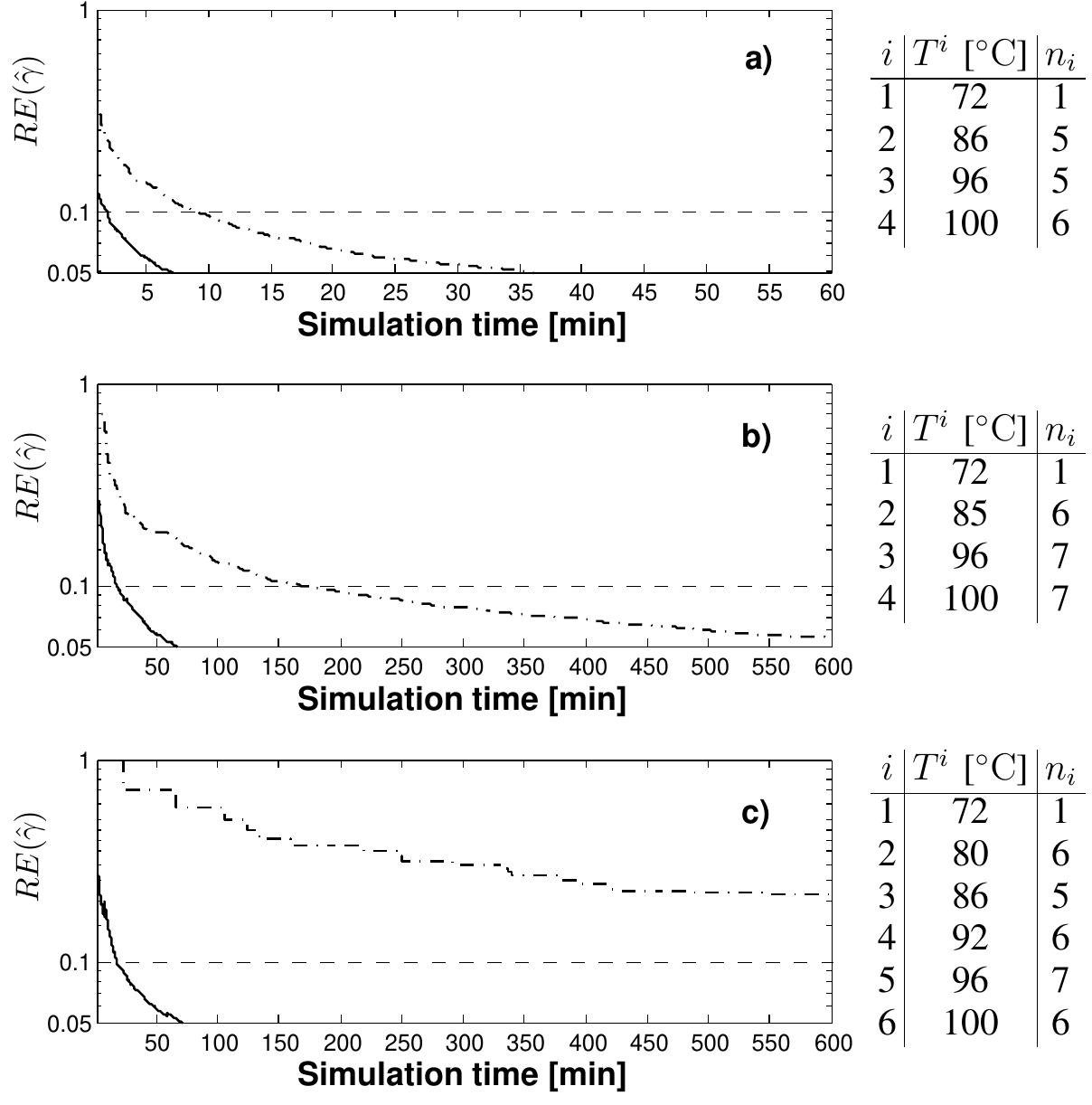}
\caption{Decrease of $RE(\hat{\gamma})$ with the simulation time. Dashed-dotted line: crude simulation, continuous line: accelerated simulation. a) $f_g=0.01$ h$^{-1}$, $C=5$. b) $f_g=0.1$ h$^{-1}$, $C=2$. c) $f_g=0.01$ h$^{-1}$, $C=2$. The tables show the selected thresholds and the number of retrials.}  
\label{RE}
\end{figure}
%%%%%Figure%%%%%
Starting with a low value of $f_g$, its increase leads to a higher probability to reach 100 \textcelsius . However, as $f_g$ is exceeding a critical value, $\hat{\gamma}$ starts to decline again. This result can be explained by the thermal inertia effects according to (\ref{eq:HBE}). While the combined power injection reaches more often higher values, the average residence time of such combined states begins to fall below the minimum time needed to heat the line up to $T_\ell^r$. A larger fluctuation magnitude is leading to a higher probability to hit $T_\ell^r$. Note that the same qualitative behavior can be shown on synthetic networks by applying a parsimonious flow model \cite{Schlapfer:2010}.    

The reduction of the simulation time by the proposed acceleration algorithm in comparison to a crude Monte Carlo simulation is shown in Fig. \ref{RE}, plotting the decreasing value of $RE(\hat{\gamma})$ in time for different values of $f_g$ and $C$. The time savings for reaching a desired accuracy level are significant, whereas the higher the probability the faster is the simulation. In case of $f_g=0.01$ h$^{-1}$ and $C=2$ ($\hat{\gamma}\approx5\cdot10^{-5}$), for example, an accuracy of $RE(\hat{\gamma})=0.1$ was reached after 1050 s. Even after 10 h of simulation time, corresponding to $N \approx 7 \cdot 10^5$ trials, this accuracy level could not be reached with the crude approach. 

The practical relevance of the fundamental findings gained by this study becomes substantiated by the following application example.
\subsection{Example B: Line Temperatures within a Transmission Network Including Wind Power}
\begin{figure}[!b]
\centering
\includegraphics[width=3.5in]{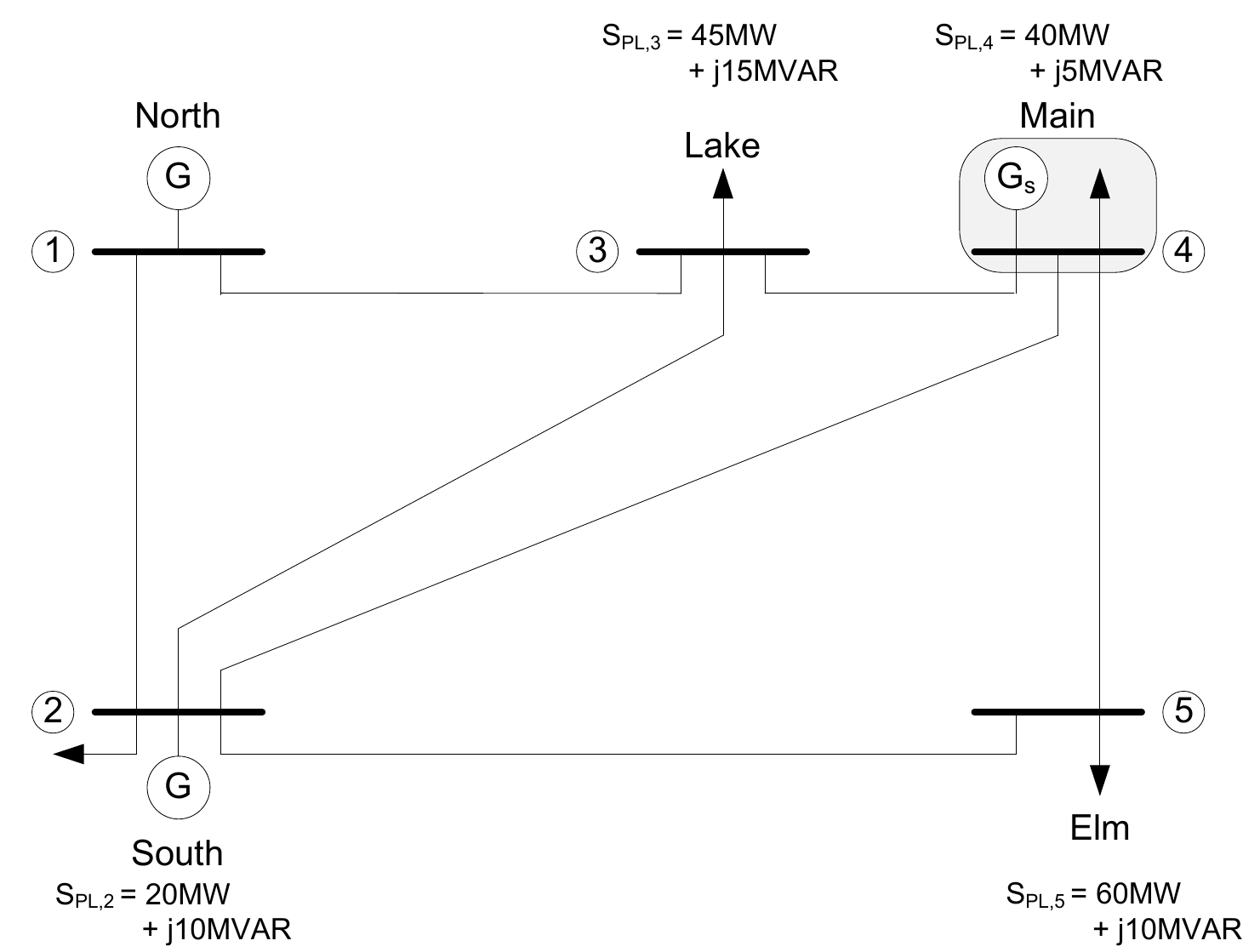}
\caption{Single line diagram of the 5-bus / 7-branch network indicating the values of the peak loads. A wind farm is connected to busbar 4.}  
\label{Fig:network}
\end{figure}

This case study demonstrates the application and practical benefits of the discussed approach for assessing the probabilities of reaching specified line temperatures within a transmission network. The exemplary 5-bus network with the electrical characteristics of the transmission lines is taken from \cite{Stagg:1968}. The additionally assigned line lengths are given in Table III in Appendix A. A single line diagram including the values of the peak loads is shown in Fig. \ref{Fig:network}. The base power is $S_B=100$ MVA.  

The fluctuation of the power flows is induced by the demand trajectory, by conventional generators being subject to random failures and by fluctuating wind power injections. A typical operational time horizon of 4 h \cite{Alguacil:2005} is chosen to be analyzed with regard to the line temperature dynamics. For the deterministic demand trajectory a typical hourly load curve is taken from \cite{Grigg:1999} (see Table IV in Appendix A).
%%%%%Figure%%%%%
%%%%%Figure%%%%%
All loads in the system follow simultaneously this curve. The total demand is covered by two conventional generating stations at busbar 1 and busbar 2 and a wind farm at busbar 4. Busbar 1 serves as the slack bus with its generating station assumed to be perfectly reliable, as generation adequacy issues are out of the scope of this paper. The second generating station at busbar 2 consists of 8 single combustion turbines with a power output of 5 MW and 3.75 MVAR each. These units are subject to independent random failures and repair processes, being conventionally modeled as a stochastic up-down-up cycle \cite{Schlapfer:2012, Billinton:1994}, corresponding to (\ref{eq:genF}). The failure and repair rates are taken from \cite{Grigg:1999} and set to $\lambda_g=1/450$ h$^{-1}$ and $\mu_g=1/50$ h$^{-1}$, respectively. The ramp rates of these generators are assumed to be sufficiently small to become neglected. The predicting time series for the fluctuating wind power generation are derived here by using a simple Markov chain model as described in \cite{Masters:2000, Mur-AmadaaIII, Papaefthymiou:2008}, with the formalism given in Appendix B. However, owing to the generality of the proposed line temperature assessment methodology, alternative probabilistic models for short-term wind power forecasting could be readily used as well, such as the Markov-switching autoregressive model \cite{Pinson:2008} or the ARIMA technique \cite{Milligan:2003, ANL:2009}, also depending on the specific meteorological conditions and data availability. The time-homogeneous and time-continuous Markov chain adopted in the present case study has been derived in \cite{Mur-AmadaaIII} for an existing wind farm. The corresponding power output states and the transition probability matrix are given in Appendix A. For all lines the `Drake' 26/7 ACSR conductor type is taken. In order to reduce complexity, all lines are assumed to be exposed to the same wind speeds in time, being perfectly correlated with the output states of the wind farm. It should be noted that a stochastic treatment of wind speed variations in both time and space is also possible in the framework of the proposed methodology. All other weather conditions are assumed to be constant and as of example A. This leads to the convected heat loss coefficients given in Table VII in Appendix A. The remaining parameters for the electrothermal model are taken from Table II. The AC power flow equations have been solved by using the Matpower \mbox{package \cite{Matpower}}. Initially, at $t_0=0$, the wind farm is in state 5, three combustion turbines are in the down state and all line temperatures are at the corresponding steady-state values. 

\mbox{Table I} shows the resulting estimates of $\gamma$ for each transmission line and different temperatures. All values have an accuracy level of $\epsilon<0.05$.
\setlength\tabcolsep{4pt}
\begin{table}[b!]
\centering
\label{tab:gammaExB}
%\vspace{-0.3cm}
\caption{Estimated probabilities $\hat{\gamma}$ of reaching the line temperature $T^i$}  
%\vspace{-0.2cm}
\begin{tabular}[c]{|c|c|c|c|c|c||c|c|}
\hline
\hline
\multirow{2}{*}{Line} & \multicolumn{7}{c|}{$\hat{\gamma}$}\\
\cline{2-8}
&  45 \textcelsius &      50 \textcelsius & 55 \textcelsius & 60 \textcelsius & 65 \textcelsius &  95 \textcelsius & 100 \textcelsius  \parbox[0pt][1.3em][c]{0cm}{} \\
\hline
1-2 & 0.80 & 0.45 &    0.27             & 0.24     &          0.12 &             0.047   &    0.0016       \parbox[0pt][1.3em][c]{0cm}{}\\
\hline
1-3 & 0.38 & 0.12   &   0.10  &    0.00029       & - & -& -   \parbox[0pt][1.3em][c]{0cm}{}\\
\hline
2-3 & 0.26 & 0.11   & - & - & - & - & -  \parbox[0pt][1.3em][c]{0cm}{}\\
\hline
2-4 & 0.26 & 0.11 & - & - & - & - & -  \parbox[0pt][1.3em][c]{0cm}{}\\
\hline
2-5 & 0.56 & 0.25  &       0.11     &            0.098  &           0.019 & - & -  \parbox[0pt][1.3em][c]{0cm}{}\\
\hline
3-4 & 0.26 & 0.097   &           - & - & - & - & -  \parbox[0pt][1.3em][c]{0cm}{}\\
\hline
4-5 & 0.24 &  0.089 & - & - & - & - & -  \parbox[0pt][1.3em][c]{0cm}{}\\
\hline
\hline
1-2, $I_{\ell,s}$& 0.80 & 0.45  & 0.27          &     0.27       &      0.12    &        0.11    &     0.061 \parbox[0pt][1.3em][c]{0cm}{} \\
\hline
\hline
\end{tabular} 
\end{table}
As both the line characteristics and weather conditions are assumed to be the same for all lines, the temperature distribution directly reflects the actual power flow distribution on the network. Furthermore, the values of $\hat{\gamma}$ for different temperatures indicate the strength of the power flow fluctuation on each line. Only line 1-2 reaches the maximum allowed operating temperature $T^r_{\ell} = 100\,^{\circ}\mathrm{C}$ with a small probability during $[t_0=0$ h, $t_e=4$ h$)$. This line becomes heavily loaded if the generating station at busbar 1 has to supply a large amount of the load in the system. This is given under the condition of a high demand level while, at the same time, the wind power output is low and several combustion turbines at busbar 2 are unavailable. The simulation time needed to compute $\hat{\gamma}$($T_{1-2}(t)=95\,^{\circ}\mathrm{C}$) with $\epsilon<0.05$ was about 3 min, for $\hat{\gamma}$($T_{1-2}(t)=100\,^{\circ}\mathrm{C}$) with $\epsilon<0.1$ about 10 min. The maximum temperatures of all other lines are found to be significantly below $T^r_{\ell}$ during the analysis period whereas the probabilities decrease sharply. These observations are referring to both a generally low loading and a small fluctuation of the power flow on these lines. 

In order to compare these probabilities as derived by the dynamic heat balance equation (\ref{eq:HBE}) with probabilistic steady-state line rating methods, we calculated the probability of reaching the current $I_{\ell,s}$, which would lead to the corresponding steady-state line temperature under the given weather conditions according to (\ref{eq:steadyState}). The resulting values are indicated in \mbox{Table I} for line 1-2. The steady-state approach significantly overestimates the probability of reaching higher temperatures. At the maximum allowed operating temperature $T^r_{\ell}=100\,^{\circ}\mathrm{C}$ the estimation of $\gamma$ differs by about a factor 40 in comparison to the proposed, more accurate approach. Indeed, at this relatively low values of $\hat{\gamma}$ the thermal inertia effect considerably reduces the probability to reach such high line temperatures within the selected time horizon, as the fluctuation frequency of the power flow exceeds the critical value identified in example A.     
%
%\vspace{0.1cm}
\section{Conclusions}
The integration of fluctuating renewable energy sources is leading to both a higher variability of the power flows and a higher operational uncertainty within the existing transmission networks. In order to adapt current loading capability assessment methods to this changing situation, this paper presents a probabilistic approach for estimating line temperatures being subject to various degrees of variability of the power flow. The temperature transients are explicitly taken into account and coupled to power flow solutions in the presence of stochastic behavior. The probabilistic approach makes use of Monte Carlo simulations and has been made computationally efficient by formulating a specific algorithm which deploys a variance reduction technique borrowed from telecommunication applications. The computational results show that the methodology allows for a fast yet accurate assessment of the operational line temperatures and associated possible overload occurrences. The presented case studies show evidence that the thermal inertia of the transmission lines can significantly reduce the probability of reaching the maximum allowed operating temperature. Due to its generality, the methodology can be combined with a broad spectrum of different probabilistic component models and therefore offers numerous applications, including a periodic operational security assessment and an accurate network capacity estimation. The practical implementation may be particularly relevant for coping with power fluctuations from large wind farms or for assessing economic transactions within a market framework. The use of online data acquired from existing SCADA systems, such as loadings, line thermal performance and actual meteorological conditions, as well as incorporating numerical weather predictions would thereby significantly increase the accuracy of the assessment.
%\vspace{0.3cm}
%
\section*{Appendix A \\ Case Study Data}
%
%\vspace{-0.1cm}
\setlength\tabcolsep{1pt}
\begin{table}[h!]
\label{tab:thermPar1}
\centering
\caption{Parameters for the electrothermal model}
%\vspace{-0.3cm}
\begin{tabular}{|c|c|c|c|c|c|c|c|cc|}
\hline \hline
$r_\ell^{ref}$         & $\alpha_\ell^{ref}$   &    $x_\ell$          &        $\rho_\ell$ &   $Q_\ell^s$ & $T_\ell^{a}$ & $T_\ell^{ref}$ & $A_\ell^c$ & $A_\ell^r$  &\\           
    $[\Omega$/m]       &     [K$^{-1}$]        &   [$\Omega$/m]       &          [J/mK]    &      [W/m]   &     [K]      &    [K]         &  [W/mK]    &  [W/mK$^4$] &\\ \hline
7.3$\cdot10^{-5}$      &  0.0039               &  2.5$\cdot10^{-4}$   &        1310        &      14.08   &  313.15      &   298.15       &   0.948    & 2.5$\cdot10^{-9}$ & \parbox[0pt][1.5em][c]{0cm}{}\\
\hline
\hline
\end{tabular}
%\vspace{-0.2cm}
\end{table}
%
%\vspace{-0.3cm}
\setlength\tabcolsep{1.4pt}
\begin{table}[h!]
\label{tab:lengths}
\centering
\caption{Length of the transmission lines} 
%\vspace{-0.3cm}
\begin{tabular}{|c|c|c|c|c|c|c|c|}
\hline \hline
Line      &   1-2  & 1-3   & 2-3  & 2-4   & 2-5   & 3-4  & 4-5 \parbox[0pt][1.5em][c]{0cm}{}\\
\hline           
Length [km] & 13.1  & 52.3  & 39.2  &  39.2  &  26.2  & 6.5  & 52.3 \parbox[0pt][1.4em][c]{0cm}{}\\
\hline
\hline
\end{tabular}
%\vspace{-0.2cm}
\end{table}
%
%
%\vspace{-0.3cm}%
\setlength\tabcolsep{2.6pt}
\begin{table}[h!]
\label{tab:load}
\centering
\caption{Hourly load levels, adopted from \cite{Grigg:1999}} 
%\vspace{-0.3cm}
\begin{tabular}{|c|c|c|c|c|}
\hline
\hline
Hour & 0$\rightarrow$1 & 1$\rightarrow$2 &  2$\rightarrow$3 & 3$\rightarrow$4  \parbox[0pt][1.4em][c]{0cm}{} \\
\hline
Load level  & 88\% &  92\% & 100\% &  97\%  \parbox[0pt][1.5em][c]{0cm}{} \\    
\hline \hline
\end{tabular}
%\vspace{-0.2cm}
\end{table}
\setlength\tabcolsep{1.4pt}
\begin{table}[h!]
\centering
\caption{Power output states of the wind farm in p.u., adopted from \cite{Mur-AmadaaIII}} 
\label{table:output}
\begin{tabular}{|c|c|c|c|c|c|c|c|c|}
\hline
\hline
&\multicolumn{8}{c|}{State}\\
\cline{2-9}
& 1 & 2 &  3 & 4 & 5 & 6 & 7 & 8 \\
\hline
$PG_w$   & 0.0112  & 0.1092  & 0.2385 & 0.4026 & 0.5409 & 0.8152 & 0.7624 & 0.9199 \parbox[0pt][1.5em][c]{0cm}{}\\
$QG_w$   & -0.0036 & -0.0061 & 0.0024 & 0.0206 & 0.0435 & 0.1012 & 0.0922 & 0.1354 \parbox[0pt][1.5em][c]{0cm}{}\\
\hline \hline
\end{tabular}
%\vspace{-0.2cm}
\end{table}
%
%\vspace{-0.3cm}
\setlength\tabcolsep{1.4pt}
\begin{table}[h!]
\label{table:pij}
\centering
\caption{Transition probability matrix $\hat{\mathbf{P}}(\nu)$ of the wind farm with $\nu=0.067$ min$^{-1}$, adopted from \cite{Mur-AmadaaIII}}  
\begin{tabular}[c]{c|c|c|c|c|c|c|c|c|c}
\multicolumn{1}{c}{} & \multicolumn{1}{c}{1} & \multicolumn{1}{c}{2}
& \multicolumn{1}{c}{3}& \multicolumn{1}{c}{4}& \multicolumn{1}{c}{5}
& \multicolumn{1}{c}{6}& \multicolumn{1}{c}{7}& \multicolumn{1}{c}{8} &\\
\cline{2-9}
1 & 0.9510 & 0.0472 & 0.0013 & 0.0001 & 0.0001 & 0.0001 & 0.0001 & 0.0001 & \parbox[0pt][1.4em][c]{0cm}{}\\
\cline{2-9}
2 & 0.1201 & 0.7643 & 0.1074 & 0.0064 & 0.0006 & 0.0006 & 0.0003 & 0.0001 & \parbox[0pt][1.4em][c]{0cm}{}\\
\cline{2-9}
3 & 0.0025 & 0.1693 & 0.7004 & 0.1156 & 0.0068 & 0.0034 & 0.0014 & 0.0006 & \parbox[0pt][1.4em][c]{0cm}{}\\
\cline{2-9}
4 & 0.0011 & 0.0057 & 0.1565 & 0.6899 & 0.1216 & 0.0181 & 0.0057 & 0.0014 & \parbox[0pt][1.4em][c]{0cm}{}\\
\cline{2-9}
5 & 0.0005 & 0.0005 & 0.0110 & 0.1760 & 0.6418 & 0.0324 & 0.1359 & 0.0019 & \parbox[0pt][1.4em][c]{0cm}{} \\
\cline{2-9}
6 & 0.0001 & 0.0015 & 0.0144 & 0.0674 & 0.1124 & 0.6116 & 0.1541 & 0.0385 & \parbox[0pt][1.4em][c]{0cm}{}\\
\cline{2-9}
7 & 0.0003 & 0.0015 & 0.0008 & 0.0051 & 0.1284 & 0.0363 & 0.6755 & 0.1521 & \parbox[0pt][1.4em][c]{0cm}{}\\
\cline{2-9}
8 & 0.0001 & 0.0001 & 0.0001 & 0.0002 & 0.0010 & 0.0044 & 0.0720 & 0.9221 & \parbox[0pt][1.4em][c]{0cm}{}\\
\cline{2-9}
\end{tabular} 
\end{table}
\begin{table}[h!]
\label{tab:Ac}
\centering
\caption{Convected heat loss coefficients for example B}  
\begin{tabular}{|c|c|c|c|c|c|c|c|c|}
\hline
\hline
Wind farm state & 1 & 2 &  3 & 4 & 5 & 6 & 7 & 8\\
\hline
$A_\ell^c$ [W/mK]  & 0.948   &  2.398 & 3.749  &  5.093  &  6.063   & 7.733  &  7.432 & 8.309  \parbox[0pt][1.5em][c]{0cm}{} \\    
\hline \hline
\end{tabular}
\end{table}
\section*{Appendix B \\ Markov Chain Formalism}
A Markov chain $X(t)$ is a random process fulfilling the (Markov) property that, given the present state, the future state is independent of the past state \cite{Grimmett:2001}:
%equation
\begin{align}\label{eq:Markov}
&Pr\big(X(t_{n+1})=j|X(t_1)=i_1,...,X(t_{n})=i_{n}\big) \nonumber\\
&=Pr\big(X(t_{n+1})=j|X(t_{n})=i_{n}\big).
\end{align}
The chain is called discrete if $X(t)$ takes values in the discrete space, and time-continuous if these values change in continuous time \cite{Grimmett:2001}. The transition probability $p_{ij}(s,t)$ is defined as
\begin{equation}
p_{ij}(s,t)=Pr\big(X(t)=j|X(s)=i\big)\qquad \textrm{for } s\leq t.
\end{equation}
If $p_{ij}(s,t)=p_{ij}(0,t-s) \equiv p_{ij}(t-s)$ for all $i,j,s,t$ the chain is said to be time-homogenous. The transition probabilities are estimated from empirical time series, whereas the states are recorded with the frequency $\nu=1/(t-s)$. The transition probabilities are then written in the stochastic matrix $\textbf{P}(\nu)$:
\begin{equation}
\mathbf{P}(\nu) = \left(
\begin{array}{cccc}
p_{11}(\nu) & p_{12}(\nu) & \ldots & p_{1n}(\nu)\\
p_{21}(\nu) & p_{22}(\nu) & \ldots & p_{2n}(\nu)\\
\vdots & \vdots & \ddots & \vdots\\
p_{n1}(\nu) & p_{n2}(\nu) & \ldots & p_{nn}(\nu)\\
\end{array} \right)
\end{equation}
where $n$ is the total number of discrete states. For each row $i$ applies $\sum_j p_{ij}(\nu)=1$. In a time-continuous Markov chain, the holding time in a given state is exponentially distributed with mean $\tau_i=1/((1-p_{ii})\nu)$.
\section*{Acknowledgment}
The authors would like to thank G\"oran Andersson (Power Systems Laboratory, ETH Zurich), Walter Sattinger (swissgrid AG) and Jos\'e Vill\'en Altamirano (Universidad Polit\'ecnica de Madrid) for their valuable suggestions. 
\bibliographystyle{IEEEtran}

\begin{thebibliography}{10}
\providecommand{\url}[1]{#1}
\csname url@samestyle\endcsname
\providecommand{\newblock}{\relax}
\providecommand{\bibinfo}[2]{#2}
\providecommand{\BIBentrySTDinterwordspacing}{\spaceskip=0pt\relax}
\providecommand{\BIBentryALTinterwordstretchfactor}{4}
\providecommand{\BIBentryALTinterwordspacing}{\spaceskip=\fontdimen2\font plus
\BIBentryALTinterwordstretchfactor\fontdimen3\font minus
  \fontdimen4\font\relax}
\providecommand{\BIBforeignlanguage}[2]{{%
\expandafter\ifx\csname l@#1\endcsname\relax
\typeout{** WARNING: IEEEtran.bst: No hyphenation pattern has been}%
\typeout{** loaded for the language `#1'. Using the pattern for}%
\typeout{** the default language instead.}%
\else
\language=\csname l@#1\endcsname
\fi
#2}}
\providecommand{\BIBdecl}{\relax}
\BIBdecl

\bibitem{Smith:2007}
J.~C. Smith, M.~R. Milligan, E.~A. DeMeo, and B. Parsons, ``Utility wind integration and operating impact state of the art,'' \emph{IEEE Trans. Power Syst.}, vol.~22, no.~3, pp. 900--908, Aug. 2007.
%
\bibitem{Banakar:2005}
H.~Banakar, N.~Alguacil, and F.~D.~Galiana, ``Electrothermal coordination part I: Theory and implementation schemes,'' \emph{IEEE Trans. Power Syst.}, vol.~20, no.~2, pp. 798--805, May 2005.
%
\bibitem{Alguacil:2005}
N.~Alguacil, M.~H.~Banakar, and F.~D.~Galiana, ``Electrothermal coordination part II: Case studies,''
\emph{IEEE Trans. Power Syst.}, vol.~20, no.~4, pp. 1738--1745, Nov. 2005.
%
\bibitem{Billinton:1996}
R. Billinton, H. Chen, and R. Ghajar, ``A sequential simulation technique for adequacy evaluation of generating systems including wind energy,'' \emph{IEEE Trans. Energy Convers.}, vol.~11, no.~4, pp. 728--734, Dec. 1996.
%
\bibitem{Schlapfer:2012}
M. Schl{\"a}pfer, T. Kessler, and W. Kr{\"o}ger, ``Reliability analysis of electric power systems using an object-oriented hybrid modeling 
approach,'' arXiv preprint arXiv:1201.0552, 2012. 
%
\bibitem{Sayas:1996}
F.~C. Sayas and R.~N. Allan, ``Generation availability assessment of
wind farms,'' \emph{Proc. Inst. Elect. Eng., Gen., Transm., Distrib.}, vol.~143, no.~5, pp. 507-–518, Sep. 1996.
%
\bibitem{Papaefthymiou:2009}
G.~Papaefthymiou and D.~Kurowicka, ``Using copulas for modeling stochastic dependence in power system uncertainty analysis,'' 
\emph{IEEE Trans. Power Syst.}, vol.~24, no.~1, pp. 40--49, Feb. 2009. 
%
\bibitem{Silva:2010}
A.~M.~L. da Silva, W.~S. Sales, L.~A. da Fonseca Manso, and R. Billinton, ``Long-term probabilistic evaluation of operating reserve requirements with renewable sources,'' \emph{IEEE Trans. Power Syst.}, vol.~25, no.~1, pp. 106--116, Feb. 2010.  
%  
\bibitem{Villen-Altamirano:2006}
M.~Vill{\'e}n-Altamirano and J.~Vill{\'e}n-Altamirano, ``On the efficiency of
  RESTART for multidimensional state systems,'' \emph{ACM Trans. Model. Comput.
  Simul.}, vol.~16, no.~3, pp. 251--279, Jul. 2006.
  
\bibitem{daSilva:2010}  
A.~M.~L. da Silva, R.~A.~G. Fernandez, and C. Singh, ``Generating capacity reliability evaluation based on Monte Carlo simulation and cross-entropy methods,'' \emph{IEEE Trans. Power Syst.}, vol.~25, no.~1, pp. 129--137, Feb. 2010
  
\bibitem{IEEE:2006}
\emph{IEEE Standard for Calculating the Current-Temperature of Bare Overhead Conductors}, IEEE Standard 738-2006, Jan. 2007.

\bibitem{Bockarjova:2007}
M.~Bockarjova and G.~Andersson, ``Transmission line conductor temperature impact on state estimation accuracy,'' in \emph{Proc. IEEE Power Tech}, Lausanne, Switzerland, Jul. 2007.

\bibitem{Deb:1968}
A.~K. Deb, \emph{Powerline Ampacity System: Theory, Modeling and
  Applications}.\hskip 1em plus 0.5em minus 0.4em\relax London: CRC Press, 2000.

\bibitem{517499}
D.~Douglass and A.-A. Edris, ``Real-time monitoring and dynamic thermal rating
  of power transmission circuits,'' \emph{IEEE Trans. Power Del.}, vol.~11, no.~3, pp. 1407--1418, Jul. 1996.
  
\bibitem{Seppa:2000}
T. O. Seppa, M. Clements, R. Payne, S. Damsgaard-Mikkelsen, and N. Coad, ``Application of real time thermal ratings for optimizing transmission line investment and operating decisions,'' in \emph{CIGRE Meeting}, Paris, France, Aug./Sep. 2000.  
  
\bibitem{Billinton:1994}
R.~Billinton and W.~Li, \emph{Reliability Assessment of Electric Power Systems
using Monte Carlo Methods}.\hskip 1em plus 0.5em minus 0.4em\relax New York: Plenum Press, 1994.

\bibitem{Milligan:2003}
M.~Milligan, M.~Schwartz, and Y.~Wan, ``Statistical wind power forecasting models: Results for U.S. wind farms,'' in \emph{Proc. Windpower 2003}, Austin, USA, May 2003.

\bibitem{ANL:2009}
\emph{Wind Power Forecasting: State-of-the-Art 2009}, Tech. Rep., Argonne National Laboratory, USA, Nov. 2009.

\bibitem{Garvels:1998}
M.~J. Garvels and D.~P. Kroese, ``A comparison of RESTART implementations,'' in
  \emph{Proc. 1998 Winter Simulation Conference}, Washington DC, USA, Dec. 1998.  

\bibitem{Matlab:2008}
Matlab, The Language of Technical Computing, Version R2008a.

\bibitem{Schlapfer:2010}
M. Schl{\"a}pfer and K. Trantopoulos, ``Lattice splitting under intermittent flows,'' \emph{Phys. Rev. E}, vol. 81, pp. 056106, 2010. 

\bibitem{Stagg:1968}
G.~W. Stagg and A.~H. El-Abiad, \emph{Computer Methods in Power System
  Analysis}.\hskip 1em plus 0.5em minus 0.4em\relax New York: McGraw-Hill, 1968.
  
\bibitem{Grigg:1999}
IEEE RTS Task Force of APM Subcommittee, ``The IEEE reliability test system - 1996,''
\emph{IEEE Trans. Power Syst.}, vol.~14, no.~3, pp. 655--667, Aug. 1999.  

\bibitem{Masters:2000}
C.~L. Masters, J. Mutale, G. Strbac, S. Curcic, and N. Jenkins, ``Statistical evaluation of voltages in distribution systems with embedded wind generation,'' \emph{
Proc. Inst. Elect. Eng., Gen., Transm., Distrib.}, vol.~147, no.~4, pp. 207--212, Jul. 2000.

\bibitem{Mur-AmadaaIII}
J.~Mur-Amada and A.~Bayod-R\'ujula, ``Wind power variability model part III -
  validation of the model,'' in \emph{Proc. 9th Int. Conf. Electrical Power Quality and Utilisation}, Barcelona, Spain, Oct. 2007.

\bibitem{Papaefthymiou:2008}
G.~Papaefthymiou and B.~Kl\"ockl, ``MCMC for wind power simulation,'' \emph{IEEE
  Trans. Energy Conversion}, vol.~23, no.~1, pp. 234--240, Mar. 2008.

\bibitem{Pinson:2008}
P. Pinson, L.~E.~A. Christensen, H. Madsen, P.~E. S\o rensen, M.~H. Donovan, and L.~E. Jensen,
``Regime-switching modelling of the fluctuations of offshore wind generation'',
\emph{J. Wind Eng. and Ind. Aerodynamics}, vol.~96, no.~12, pp. 2327--2347, Dec. 2008.

\bibitem{Matpower}
R.~D. Zimmerman, C.~E. Murillo-S\'anchez, and R. J. Thomas, ``Matpower's extensible optimal power flow architecture,'' in \emph{Proc. IEEE Power and Energy Society General Meeting}, Calgary, Canada,
Jul. 2009.

\bibitem{Grimmett:2001}
G.~R. Grimmett and D.~R. Stirzaker, \emph{Probability and Random
  Processes}.\hskip 1em plus 0.5em minus 0.4em\relax {Oxford: Oxford University Press},
  2001.
%
\end{thebibliography}

%

%
%\vfill
%
\end{document}